\documentclass[lettersize,journal]{IEEEtran}
\usepackage{amsmath,amsfonts}
\usepackage{algorithmic}
\usepackage{algorithm}
\usepackage{array}
\usepackage[caption=false,font=normalsize,labelfont=sf,textfont=sf]{subfig}
\usepackage{textcomp}
\usepackage{stfloats}
\usepackage{url}
\usepackage{verbatim}

\usepackage{graphicx}
\usepackage{cite}
\usepackage{amsmath}
\usepackage{amssymb}
\usepackage{amsfonts}
\usepackage{amssymb}
\usepackage{amsthm}
\usepackage{graphicx}
\usepackage{booktabs}
\usepackage{algorithm}
\usepackage{algorithmic}
\usepackage{flushend}

\usepackage{amssymb}
\usepackage{amsmath,amsfonts}
\usepackage{algorithmic}
\usepackage{array}
\usepackage{textcomp}
\usepackage{stfloats}
\usepackage{url}
\usepackage{verbatim}
\usepackage{graphicx}
\usepackage{hyperref}
\usepackage{fontawesome5}
\usepackage{amsthm}
\usepackage{xcolor}
\newtheorem{theorem}{Theorem}

\newtheorem{identity}[theorem]{Identity}
\theoremstyle{remark}
\newtheorem*{remark}{\textbf{Remark}}

\begin{document}

\title{Online/Offline Learning to Enable Robust Beamforming: Limited Feedback Meets Deep Generative Models}

\author{Ying Li, Zhidi Lin, Kai Li, and Michael Minyi Zhang
\thanks{ Ying Li and Michael Minyi Zhang are with the Department of Statistics and Actuarial Science, University of Hong Kong, Hong Kong, SAR, China (e-mail: lynnli98@connect.hku.hk; mzhang18@hku.hk). 

Zhidi Lin and Kai Li are with the School of Science and Engineering, Chinese University of Hong Kong (Shenzhen), Shenzhen 518172, China, and also with Shenzhen Research Institute of Big Data, Shenzhen 518172, China (e-mail:zhidilin@link.cuhk.edu.cn; kaili4@link.cuhk.edu.cn).
}
}

\markboth{Journal of \LaTeX\ Class Files,~Vol.~14, No.~8, August~2021}%
{Shell \MakeLowercase{\textit{et al.}}: A Sample Article Using IEEEtran.cls for IEEE Journals}


\maketitle

\begin{abstract}
Robust beamforming is a pivotal technique in massive multiple-input multiple-output (MIMO) systems as it mitigates interference among user equipment (UE). One current risk-neutral approach to robust beamforming is the stochastic weighted minimum mean square error method (WMMSE). However, this method necessitates statistical channel information, which is typically inaccessible, particularly in fifth-generation new radio frequency division duplex cellular systems with limited feedback. To tackle this challenge, we propose a novel approach that leverages a channel variational auto-encoder (CVAE) to simulate channel behaviors using limited feedback, eliminating the need for specific distribution assumptions present in existing methods. To seamlessly integrate model learning into practical wireless communication systems, this paper introduces two learning strategies to prepare the CVAE model for practical deployment. Firstly, motivated by the digital twin technology, we advocate employing a high-performance channel simulator to generate training data, enabling pretraining of the proposed CVAE while ensuring non-disruption to the practical wireless communication system. Moreover, we present an alternative online method for CVAE learning, where online training data is sourced based on channel estimations using Type II codebook. Numerical results demonstrate the effectiveness of these strategies, highlighting their exceptional performance in channel generation and robust beamforming applications.
\end{abstract}


\section{Introduction}

\IEEEPARstart{T}{he} development of the fifth-generation (5G) cellular systems \cite{5g-1, 5g-2, sevgican2020intelligent} and beyond, such as 6G \cite{ jiang2021road, garcia2021tutorial, kim2023towards}, faces several challenges. These challenges include temporally and spatially varying wireless communication environments, limited availability of the radio frequency spectrum, and growing demand for higher network capacity, data rate, and quality of service. Recently, massive multiple-input multiple-output (MIMO) systems \cite{rapajic2000information, truong2013effects, shi2023channel} have been widely utilized in the telecommunication industry as one of the most crucial technologies solving these aforementioned pressing issues. Specifically, massive MIMO is a radio antenna technology that deploys many antennas at both base station (BS) and user equipment (UE). The key concept in massive MIMO is to apply spatial multiplexing and beamforming for transmitting independent and distinct directionally encoded data signals, known as ``streams'', thus enabling the reuse of the same time and frequency resources.

Specifically, the beamforming technique \cite{gershman2010convex, huang2019fast} precodes the signal, modifying its phase and amplitude in baseband processing before radio frequency (RF) transmission. This modification allows the signal to be focused on a specific direction rather than being broadcast over a wide area. As a result, the beamforming technique can mitigate interference from multiple UE (multi-UE) and significantly improve system capacity \cite{huang2019fast}.

The efficacy of the beamforming technique hinges on the availability of precise channel state information (CSI) at the BS \cite{cheng2021towards}. In time division duplex (TDD) systems, the BS and UE share the same frequency resource, which implies that CSI measurements at the downlink and the uplink channel are equally informative. Thus, it is convenient to use the uplink CSI as an approximation of the downlink CSI. However, this scheme is impossible in practical 5G new radio frequency division duplex (5G NR FDD) cellular systems due to the lack of the aforementioned channel reciprocity \cite{love2008overview, guey2004modeling}. 

To tackle this challenge, recent work has utilized a limited feedback scheme in practical 5G NR FDD cellular systems \cite{code1} . Specifically, a set of vectors (known as the codebook) is shared at both the BS and the UE for the purpose of CSI quantification.\footnote{In this paper, the Type I \cite{code1} and Type II \cite{liu2016impact} codebooks are employed, in accordance with the specifications outlined in the 5G NR standards \cite{R16, R17}.} In each time slot, the UE estimates the CSI and subsequently provides feedback to the BS in the form of a precoder matrix indicator (PMI) and a channel quality indicator (CQI). The PMI corresponds to the index of the vector closest to the CSI, while the CQI indicates the channel quality. Due to the existence of quantification errors and CSI estimation errors, the representation capability of feedback is rather limited. Therefore, it is crucial to develop an advanced beamforming technique that is robust to both CSI estimation and quantification errors. 

\subsection{Related Works}
Most existing work on beamforming centers around the development of methods that rely on accurate CSI \cite{wmmse}. For example, the iterative weighted minimum mean square error (WMMSE) method \cite{wmmse} is widely employed as a benchmark and serves as the foundation for robust beamforming approaches. Rather than requiring perfect and complete CSI, the stochastic WMMSE method introduced in \cite{ssum} offers an alternative approach by relying on statistical channel knowledge (or knowledge of the channel statistical distribution). This method optimizes the average performance using the stochastic sequential upper-bound minimization (SSUM) framework \cite{ssum}. However, this assumption is questionable: \emph{How can we obtain the statistical channel knowledge or corresponding channel samples with minimal feedback information (e.g., PMIs and CQIs) in practical 5G NR FDD systems?} 

Several attempts have been made to model the channel statistical distribution based on feedback. Specifically, \cite{1, 2, 3} assume statistical independence across both UE and spatial dimensions and use a multivariate Gaussian distribution with a diagonal covariance matrix as the channel statistical distribution, with the mean estimated from the PMI. However, these assumptions are not always satisfied in practice \cite{real-channel-1, real-channel-2, real-channel-4}. A more realistic channel model \cite{Gauss-markov-1} employs a jointly correlated Gaussian distribution to model the spatial correlations across the antennas. Nonetheless, the Gaussian assumption  might still significantly deviate from practical channel measurements \cite{real-channel-1}.  Instead of assuming the simplistic Gaussian model or other specific statistical parametric models  \cite{1, 2, 3, Gauss-markov-1},  this paper turns to deep generative models (DGMs) that have superior modeling capacity for complex distributions \cite{theodoridis2020machine}. 

The most popular DGMs include variational auto-encoders (VAEs) \cite{vae, cvae, li2023overcoming}, generative adversarial networks (GANs) \cite{goodfellow2020generative}, normalization flows \cite{kobyzev2020normalizing}, energy-based models \cite{yu2020training}, and diffusion models \cite{croitoru2023diffusion}, to name a few.  In our research context, our intent is to model the channel statistical distribution so that the generated channel vectors are pair-wise close to the ground-truth channel vectors in terms of cosine similarity \cite{liang2020deep}. However, 
incorporating this objective into the DGMs (like in the GANs framework) is not straightforward \cite{goodfellow2020generative}. This paper centers its focus on VAEs \cite{vae}, a fundamental framework that is comprised of two essential components: an encoder and a decoder. The encoder is responsible for mapping high-dimensional input data into a lower-dimensional latent space, capturing the underlying essential features and variations present in the data; on the other hand, the decoder faithfully reconstructs the data by projecting samples from the latent space back into the original high-dimensional space. VAEs have demonstrated remarkable generative capabilities in a diverse range of domains, including computer vision \cite{razavi2019generating}, natural language processing \cite{zhang2022effect}, and audio synthesis \cite{li2021audio2gestures}. Subsequent sections of this paper will elucidate the incorporation of cosine similarity into the reconstruction aspect of VAE, accomplished through the implicit noise assumption. 

\subsection{Contributions}
To approach the unknown ground-truth channel statistical distribution, this paper devises a VAE-based channel generating system, called channel VAE (CVAE). This system can mimic the channel generative behaviors of the practical wireless environment within the context of 5G NR FDD massive MIMO cellular systems using Type I codebook \cite{code1}. The major contributions of this paper are summarized as follows:
\begin{itemize}
    \item In contrast to previous works that rely on a particular statistical distribution, we propose a system \cite{1, 2, 3, Gauss-markov-1}, called CVAE, tailored for simulating the channel statistical distribution given the limited feedback without making specific assumptions about the distribution. However, it is crucial to note that this model necessitates training with data before its practical deployment. This raises the question that how to seamlessly integrate model learning into practical wireless communication environments. In this paper, we present two learning methods before the system deployment. 
    \item 
    For the first approach, we propose to train the CVAE system in an offline fashion using a high-performance channel simulator, e.g., QUAsi Deterministic RadIo channel GenerAtor (QuaDRiGa) \cite{real-channel-1}. This approach is inspired by the recent concept of a digital twin \cite{fuller2020digital}. Specifically, we utilize the channel simulator to generate an extensive level of training data, closely mirroring authentic channel characteristics. Then, the data are used for fully training the CVAE model with substantial parameters. Notably, this process occurs prior to online deployment, guaranteeing no impact on the operation of practical systems. 
    \item For the second approach, in case the high-performance channel simulator is unavailable, we propose a Type II codebook-assisted online learning scheme, in which Type II codebook-based channel estimates are used for training, and a separate CVAE is trained for each UE. The online learning scheme can be more adaptive to the fast-changing wireless environments and provide robust beamforming for each UE. 
    \item The channel statistical distributions generated by those two approaches enable the practical implementation of the stochastic WMMSE technique in 5G NR FDD cellular systems. Our numerical results demonstrate the efficacy of the proposed CVAE and the associated learning methods in enhancing the performance of robust beamforming.
\end{itemize}
A portion of the research in this paper was previously presented in IEEE SPAWC 2021 \cite{li2021digital}. However, this paper offers additional contributions and a more comprehensive presentation, encompassing a solution for the online scenario and supplementary simulation results not included in the conference version \cite{li2021digital}.

\subsection{Paper Organization}
The remainder of this paper is organized as follows. In Section \ref{sec:system_model_roubst_beamforming}, the system model and the basic idea of the stochastic WMMSE algorithm are introduced.  Section \ref{sec:existing_solutions} presents existing solutions that rely on Type I codebook-based channel estimates. In Sections \ref{sec:digital_twin_inspired_offline_learning} and \ref{sec:online_learning}, the digital twin-inspired offline training scheme and Type II codebook-assisted online training scheme are introduced, respectively.  Based on those approaches, numerical results and discussions are presented in Section \ref{sec:experiments}. Finally, conclusions are drawn in Section \ref{sec:conclusion}.

\section{System Model and Stochastic WMMSE} \label{sec:system_model_roubst_beamforming}
This section presents the preliminary concepts essential to understanding this paper. Specifically,
Section \ref{subsec:system} introduces the system model, and Section \ref{subsec:stochastic_wmmse} summarizes the basic idea of stochastic WMMSE \cite{ssum}. Finally, the necessity of robust beamforming based on stochastic WMMSE is highlighted in Section \ref{subsec:Why_sto_wmmse}. 

\subsection{System Model}
\label{subsec:system}
In the context of a single-cell multi-UE broadcast channel, it should be noted that the BS with $N_A$ transmit antennas is capable of serving $L$ single-antenna UEs. The $i$-th UE is denoted as UE $i$, and the downlink channel vector between the BS and UE $i$ is denoted as $\boldsymbol{h}_i \in \mathbb{C}^{N_A}$. 
We assume that the channel vector $\boldsymbol{h}_i$ remains invariant within one time slot but may change afterward. To simplify implementation, linear precoding is utilized to transmit streams $\left\{ s_{i} \in \mathbb{C} \right\}_{i=1}^{L}$, where streams are independent Gaussian random variables, and it holds that $\mathbb{E}[ s_{i} \bar{s}_{i} ] = 1, \forall i $ \cite{ssum}.  Furthermore, at the BS, the transmitted signal $\boldsymbol{x}$ is 
\begin{align}
\boldsymbol x = \sum_{i=1}^{L} \boldsymbol v_{i} s_{i},
\end{align}
where $\boldsymbol v_{i} \in \mathbb C^{N_A}$ represents the beamforming vector of UE $i$. Due to the power consumption restriction at the BS, the average transmission power is constrained by a budget of $P$, i.e.,
\begin{align}
\sum_{i=1}^{L} \operatorname{Tr}\left(\boldsymbol{v}_{i} \boldsymbol{v}_{i}^{H}\right) \leq P.
\end{align}
Transmitting through the channel, the received data at UE $i$ is
\begin{align}
\label{eq:received_signal}
    y_i = \boldsymbol{h}_i^{H} \boldsymbol{x} + n_i  
    = \boldsymbol{h}_i^{H} \boldsymbol{v}_i s_i + \sum_{i \neq l} 
    \boldsymbol{h}_i^{H} \boldsymbol{v}_l s_l + n_i , ~ \forall i, 
\end{align}
where ${\boldsymbol{h}_i^{H} \boldsymbol{v}_i s_i}$ and ${\sum_{i \neq l} \boldsymbol{h}_i^{H} \boldsymbol{v}_l s_l}$ are the desired signal of $i$-th UE and the multi-UE interference, respectively; $n_{i}$ represents the additive white Gaussian noise (AWGN) with distribution $\mathcal {CN}(0, \sigma_{i}^2)$. Here $\sigma_i$ serves as the hyperparameter representing uncertainty. We assume that the signals transmitted to various UEs are statistically independent \cite{wmmse}. Then, treating interference as noise, a linear detector $ u_{i} \in \mathbb C$ is used at each UE $i$ to estimate the data stream $s_{i}$, i.e.,
\begin{align}
\hat{s}_{i}=\bar{u}_{i} y_{i}, ~ \forall i . 
\end{align}

\subsection{Stochastic WMMSE for Robust Beamforming}
\label{subsec:stochastic_wmmse}

Assuming that the channel statistical distribution $p(\boldsymbol h), \boldsymbol h \triangleq \{ \boldsymbol h_i \}_{i=1}^{L}$ is accessible at the BS, the objective of robust beamforming, as discussed in \cite{ssum}, is to find optimal beamforming vectors that maximize the average sum-rate of all UEs, expressed mathematically as
\begin{align}
\begin{aligned}
  & \mathrm{max}_{\boldsymbol{v}} \ \mathbb{E}_{\boldsymbol{h}}\left[ \sum_{i=1}^{L} \max _{u_{i}} \left\{R_{i}\left(u_{i}, \boldsymbol{v}, \boldsymbol{h}\right)\right\}\right],  \\ 
  & \text { s.t. }  \sum_{i=1}^{L} \operatorname{Tr}\left(\boldsymbol{v}_{i} \boldsymbol{v}_{i}^{H}\right) \leq P . 
\end{aligned}
\label{ssum_rate}
\end{align}
Here the instantaneous achievable rate $R_{i}\left(u_{i}, \boldsymbol{v}, \boldsymbol{h}\right)$ of UE $i$ is defined as
\begin{align}
R_{i}\left({u}_{i}, \boldsymbol{v}, \boldsymbol{h}\right) \triangleq \log \operatorname{det}\left(\boldsymbol{E}_{i}^{-1}\left( u_{i}, \boldsymbol{v}, \boldsymbol{h}_i \right)\right), 
\end{align}
where 
\begin{align} 
\boldsymbol{E}_{i}\left(u_{i}, \boldsymbol{v}, \boldsymbol h_i \right) \triangleq &\left(1-\bar{u}_{i} \boldsymbol{h}_{i}^H \boldsymbol{v}_{i}\right)\left(1-\bar{u}_{i} \boldsymbol{h}_{i}^{H} \boldsymbol{v}_{i}\right)^{H} \\&+ \sum_{l \neq i} \bar{u}_{i} \boldsymbol{h}_{i}^H \boldsymbol{v}_{l} \boldsymbol{v}_{l}^{H} \boldsymbol{h}_{i} u_{i}+\sigma_{i}^{2} \bar{u}_{i} u_{i} \nonumber
\end{align}
represents the mean square error (MSE) matrix \cite{ssum}.

It is evident that the optimal linear detector $u_i$, which maximizes the problem defined in \eqref{ssum_rate}, corresponds to the minimum MSE (MMSE) detector \cite{wmmse}, 
\begin{align}
    u_{i}^{\mathrm{MMSE}}= \frac{1}{J_i} \boldsymbol{h}_i^{H} \boldsymbol{v}_i , \label{update_u}
\end{align}
where $J_i \triangleq  \sum_{l=1}^{L} \boldsymbol{h}_{i}^{H} \boldsymbol{v}_{l} \boldsymbol{v}_{l}^{H} \boldsymbol{h}_{i} +\sigma_{i}^{2}$. 

To solve the optimization problem formulated in \eqref{ssum_rate}, a conventional and effective method employed is the sample average approximation (SAA) technique. More specifically, during iteration $r$ of this method, a new realization of the channel vector $\boldsymbol{h}^{r} \triangleq \{ \boldsymbol{h}^{r}_i \}_{i=1}^{L}$ is acquired, and the beamforming vectors are updated by solving 
\begin{equation}
    \begin{aligned}
    & \min _{\boldsymbol{v}} \ \frac{1}{r} \sum_{k=1}^{r} g(\boldsymbol{v}, \boldsymbol{h}^{k}), \\  
    & \text { s.t. } \sum_{i=1}^{L} \operatorname{Tr}\left(\boldsymbol{v}_{i} \boldsymbol{v}_{i}^{H}\right) \leq P.  
\end{aligned}
\label{eq:saa}
\end{equation}
where 
$
    g(\boldsymbol{v}, \boldsymbol{h}^{k})= \sum_{i=1}^{L} -R_{i}\left(u_{i}^{\text{MMSE}}, \boldsymbol{v}, \boldsymbol{h}^{k}\right). \label{eq:opt_u}
$
Nonetheless, a significant drawback associated with the SAA method lies in the complexity of each of its steps. In general, solving \eqref{eq:saa} is challenging due to the non-convex nature of the objective function,  $\frac{1}{r} \sum_{k=1}^r g(\boldsymbol{v}, \boldsymbol{h}^{k})$. 
To address this challenge, the SSUM framework proposed in \cite{ssum}, offers a solution. Instead of directly optimizing the non-convex objective function $\frac{1}{r} \sum_{k=1}^r g(\boldsymbol{v}, \boldsymbol{h}^{k})$,  SSUM minimizes a locally tight convex upper bound of this function.  

To determine this locally tight convex upper bound,  a set of auxiliary variables $\boldsymbol{p} \triangleq(w, \boldsymbol{z}) = (\{w_i\}_{i=1}^L, \{\boldsymbol{z}_i\}_{i=1}^L)$ is firstly defined \cite{ssum}, where $w_{i} \in \mathbb{C} $
and $\boldsymbol{z}_i \in \mathbb{C}^{N_A}$. Then, for some fixed $\rho > 0$, we let 
\begin{align}
        \hat{R}_{i}\left(w_{i}, \boldsymbol{z}_{i}, \boldsymbol{v}, \boldsymbol{h}^{k}\right) \triangleq &-\log \operatorname{det}\left(w_{i}\right)+\operatorname{Tr}\left(w_{i} \boldsymbol{E}_{i}\left( u_i^{\text{MMSE}}, \boldsymbol{v}, \boldsymbol h^{k}\right)\right) \nonumber \\ &+  \rho\left\|\boldsymbol{v}_{i}-\boldsymbol{z}_{i}\right\|^{2}-d_{i}, 
\end{align}
and define
\begin{align}
    \mathcal{G}(\boldsymbol{v}, \boldsymbol{p}, \boldsymbol{h}^{k}) \triangleq \sum_{i=1}^{L} \hat{R}_{i}\left(w_{i}, \boldsymbol{z}_{i}, \boldsymbol{v}, \boldsymbol{h}^{k}\right).
\end{align}
It can be shown that the constructed function, $\mathcal{G}(\boldsymbol{v}, \boldsymbol{p}, \boldsymbol{h}^{k})$, is as a locally tight convex upper bound of the function, $g(\boldsymbol{v}, \boldsymbol{h}^{k})$, as discussed in \cite{ssum}.  Subsequently, at each iteration $r$, SSUM comprises two alternating steps. The first step involves finding the optimal $\boldsymbol{p}^{r}$ by solving
\begin{align}
    \boldsymbol{p}^{r} \triangleq \arg \min _{\boldsymbol{p}} \mathcal{G}\left(\boldsymbol{v}^{r-1}, \boldsymbol{p}, \boldsymbol{h}^{r}\right). \label{ssump}
\end{align}
This step aims to derive a locally tight convex upper bound for the function $g(\boldsymbol{v}, \boldsymbol{h}^{r})$ at $\boldsymbol{v}^{r-1}$, where $\boldsymbol{v}^{r-1}$ represents beamforming vectors at the $(r-1)$-th iteration. Upon scrutinizing the first-order optimality condition of \eqref{ssump}, it becomes apparent that updates for the variable $\boldsymbol{p}^{r} = (w, \boldsymbol{z})$ can be obtained analytically.
The other step involves updating the beamforming vectors by replacing the non-convex function $g(\boldsymbol{v}, \boldsymbol{h}^{k}), \forall k \in 1, ..., r$ in \eqref{eq:saa} with its locally tight convex upper bound, which can be mathematically expressed as:
\begin{align}
    \begin{aligned} 
    \boldsymbol{v}^{r} = & \arg \min_{\boldsymbol{v}} \frac{1}{r} \sum_{k=1}^{r} \mathcal{G} \left(\boldsymbol{v}, \boldsymbol{p}^{k}, \boldsymbol{h}^{k}\right) \\ 
    & \text { s.t. } \sum_{i=1}^{L} \operatorname{Tr}\left(\boldsymbol{v}_{i} \boldsymbol{v}_{i}^{H}\right) \leq P.\end{aligned}
    \label{problem_v}
\end{align}
Utilizing the Lagrange function for the problem \eqref{problem_v}, the first-order optimality condition with respect to each $\boldsymbol{v}_i$ provides a closed-form solution with a Lagrangian multiplier, which can be determined through a one-dimensional search \cite{wmmse}.

\subsection{The Necessity of Robust Beamforming}
\label{subsec:Why_sto_wmmse}
Practical communication systems are prone to various errors, such as channel estimation and quantization errors, among others. Consequently, the CSI received at the BS may not match the true CSI. Therefore, instead of assuming the exact CSI, it is beneficial to consider the corresponding channel statistical distribution, as it enables fault tolerance and improves robustness \cite{ssum}. 

To illustrate the superior performance of stochastic WMMSE \cite{ssum} in robust beamforming, we provide an example using statistical channel knowledge. We generate $100$ channel samples for $4$ UEs using QuaDRiGa \cite{real-channel-1} and create a complex Gaussian distribution for each of these samples. This distribution, denoted as $p(\boldsymbol{h})$, has the true channel vector as its mean and an identity matrix scaled by the variance of noise as its covariance matrix.

\begin{figure}[!t]
    \centering
    \includegraphics[width=2.3in]{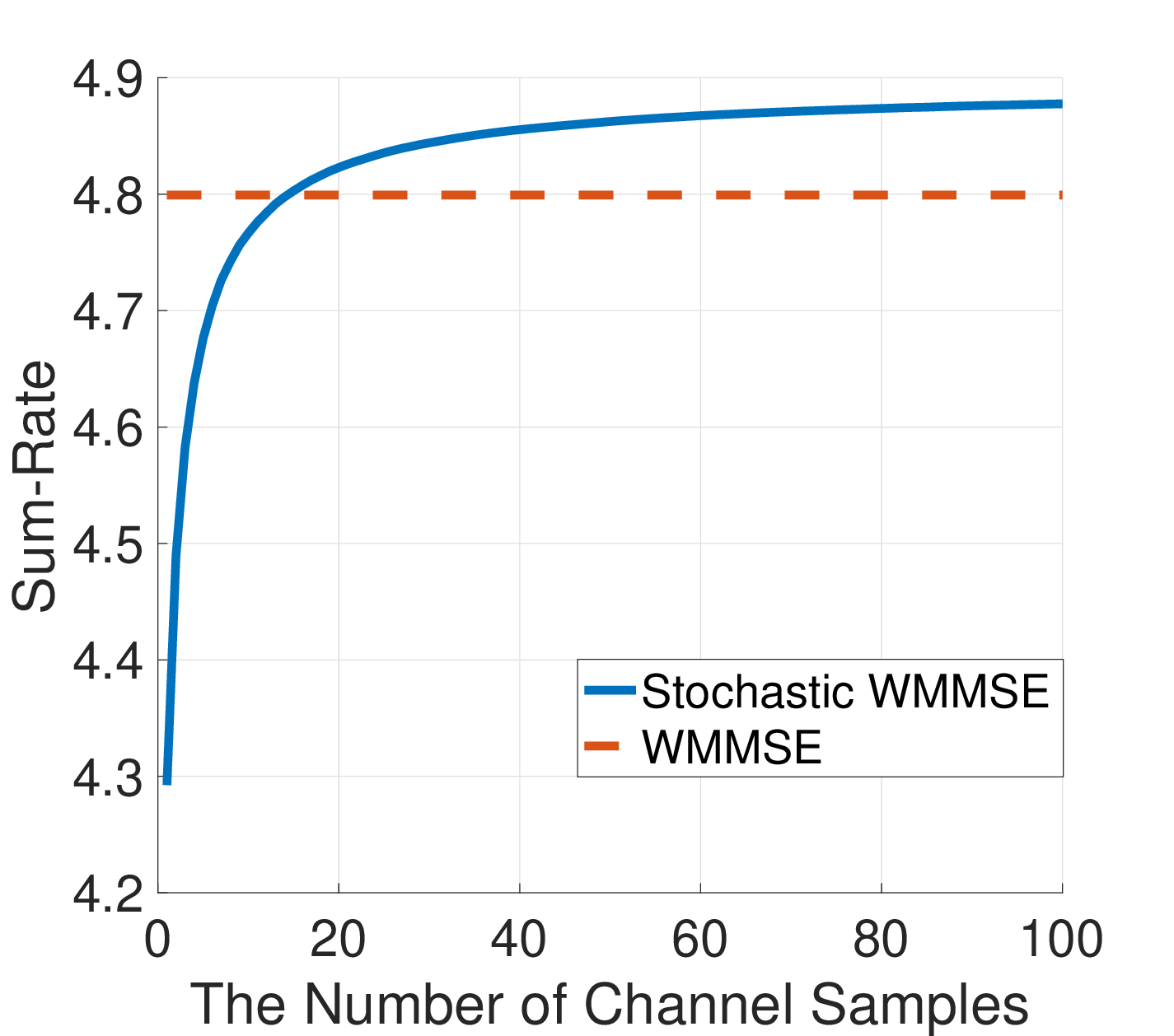}
    \caption{The sum-rates from stochastic WMMSE and WMMSE versus the number of channel samples.}
    \label{fig:motivation}
 \end{figure}

We draw $100$ channel samples from this distribution and calculate the associated sample mean $\tilde{\boldsymbol h}$. These $100$ samples are used as input for stochastic WMMSE, while the mean $\tilde{\boldsymbol h}$ serves as input for the conventional WMMSE. Fig. \ref{fig:motivation} presents a comparison of the performance of two algorithms in terms of the sum-rate (defined in \eqref{ssum_rate}) as the number of channel samples varies. The data presented is an average of 100 Monte-Carlo trials, each using a ground-truth channel sample. It is evident that the stochastic WMMSE algorithm outperforms the conventional WMMSE, surpassing it after processing only $17$ samples.

However, the promise offered by the stochastic WMMSE algorithm heavily relies on channel samples from the channel statistical distribution $p(\boldsymbol{h})$, which is challenging to acquire in practice. Therefore, to enable the practical use of stochastic WMMSE, a natural question arises: \emph{In practical 5G NR FDD cellular systems, how can we obtain channel samples from their channel statistical distribution?} 

\section{Existing Solutions: Estimating Channels from Limited Feedback} \label{sec:existing_solutions}

Before presenting our channel sample acquisition solutions, we will first review existing methods for channel estimation in practical 5G NR FDD cellular systems with limited feedback. The principle of the existing solutions is to estimate channel vectors through a Type I codebook-based feedback. Specifically, a codebook $\mathcal{A}=\{ \mathbf{a}_m \in \mathbb{C}^{N_P} , m=1, \ldots, M \}$, such as Type I codebook \cite{code1}, is shared by the BS and all UEs. Here, $N_P$ represents the number of ports. In each communication round, the dimension of the pilot signal $\boldsymbol{r}$ is first enlarged by the virtual antenna matrix $\boldsymbol{Q} \in \mathbb{C}^{N_A \times N_P}$ (also known as the pilot weighting matrix \cite{q}), i.e., $\boldsymbol{Q}\boldsymbol{r} \in \mathbb{C}^{N_A}$, before being transmitted to each UE over downlink channels.

As a result, the data received at each UE is expressed as
\begin{align}
    y_{i} = \boldsymbol h_i^{H} \boldsymbol Q \boldsymbol r + z_i, ~ \forall i, 
\end{align}
where $z_i \sim \mathcal{CN}(0, \sigma_i)$ represents the AWGN. Since the virtual antenna matrix is known, the effective channel covariance matrix can be calculated at the UE: 
\begin{align}
    \boldsymbol{R}_i \triangleq  \mathbb{E}\left[ \boldsymbol{Q}^H \boldsymbol{h}_i \boldsymbol{h}_i^{H} \boldsymbol{Q} \right]= \boldsymbol{Q}^{H} \boldsymbol{C}_{i} \boldsymbol{Q}, 
\end{align}
where $\boldsymbol{C}_{i} \triangleq \mathbb{E}\left[\boldsymbol{h}_{i} \boldsymbol{h}_{i}^{H}\right]$. 
Using the effective channel covariance matrix $\boldsymbol{R}_i$, the PMI
\begin{align}
    m_i = \arg \max_{m=1, \ldots, M} \boldsymbol {a}_{m}^{H} \boldsymbol{R}_i \boldsymbol{a}_{m}
\end{align}
and CQI 
\begin{align}
    \eta_i = \max_{m=1, \ldots, M} \boldsymbol {a}_{m}^{H} \boldsymbol{R}_i \boldsymbol{a}_{m}
\end{align}
are calculated and then fed back to the BS.

Once PMIs $\{ m_i\}_{i=1}^L$ have been transmitted by all UEs, the vector $\hat{\boldsymbol h}_i \triangleq \boldsymbol Q \boldsymbol a_{m_i}, ~ \forall i \in 1, ..., L$, 
is regarded as the downlink channel estimate \cite{1, Gauss-markov-1} at the BS. 
These estimations can be utilized to construct an empirical distribution of CSI, denoted as $\left\{\hat{\boldsymbol{h}}_{i}\right\}_{i=1}^{L}$, i.e., 
\begin{align}
    p(\boldsymbol h \mid m) = \frac{1}{L} \sum_{i=1}^{L} ( \delta_{\hat{\boldsymbol h}_i} + n_i), ~ \text{where} ~ \delta_{\hat{\boldsymbol h}_i} = \left\{ \begin{array}{c}
        1 ~~ \text{if} ~~ \boldsymbol{h} = \boldsymbol{h}_i\\
        0 ~~ \text{otherwise} 
   \end{array} \right. , 
\end{align}
which can be employed as the input of the stochastic WMMSE algorithm. Recall that, $n_{i}$ represents the AWGN with distribution $\mathcal {CN}(0, \sigma_{i}^2)$. 

The current Type I codebook-based approach is simple but exhibits limited beamforming performance. This limitation arises from the use of a limited number of Type I code-vectors \cite{code1} for communication efficiency. As a result, it leads to coarse channel estimations $\{ \hat{\boldsymbol{h}}_i \}_{i=1}^{L}$. Therefore, it is essential to develop advanced techniques capable of generating accurate channel samples from limited feedback information, such as PMIs and CQIs. In this paper, we illustrate how this objective can be accomplished through the proposed CVAE, which can operate in offline or online modes for various purposes. Further details on these two proposed schemes are provided in the subsequent sections.

\begin{figure*}[!t]
    \centering
    \includegraphics[width= 0.8 \linewidth]{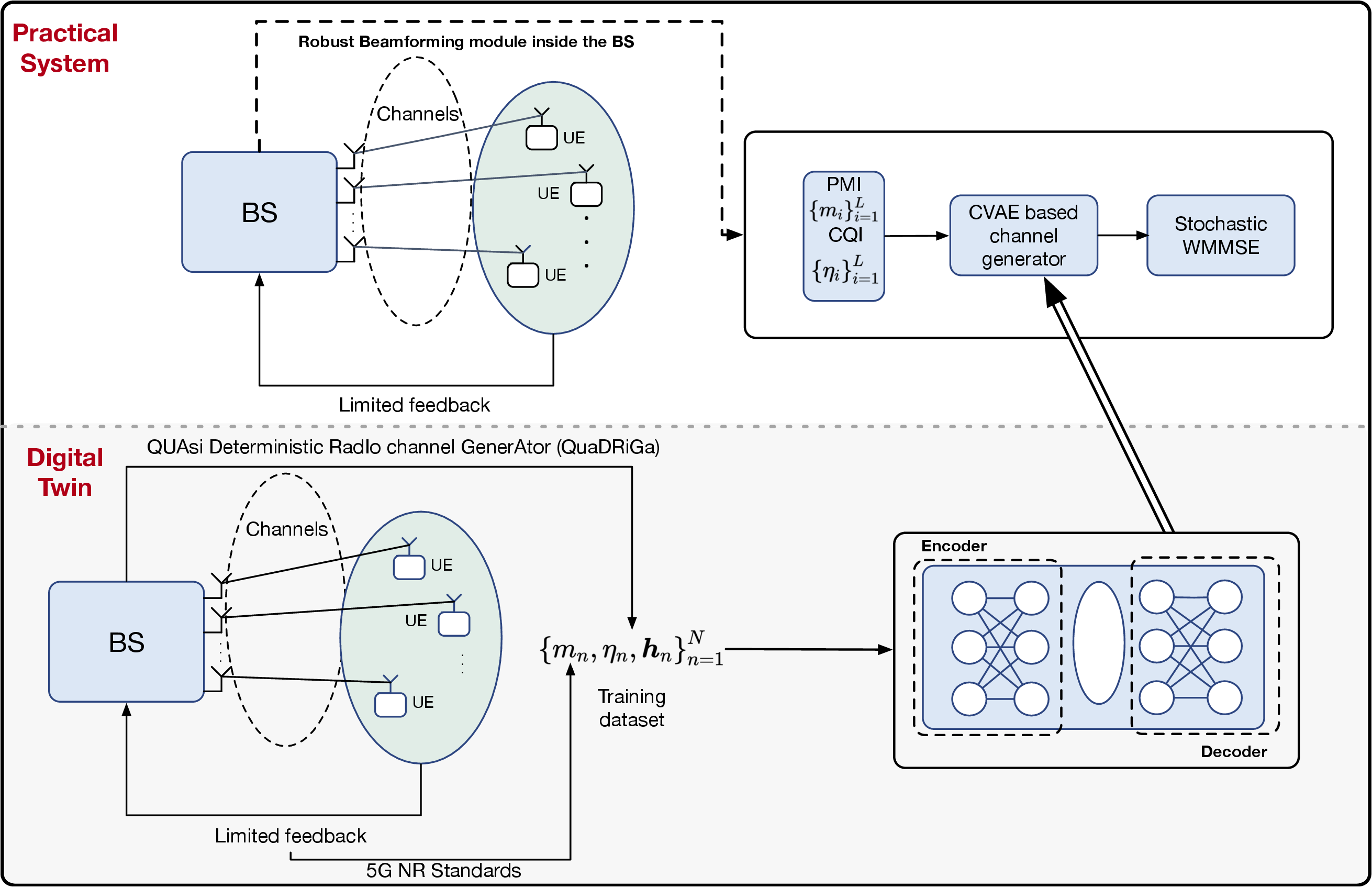}
    \caption{The proposed digital twin-inspired offline learning scheme. The digital twin system, which is based on the QuaDRiGa channel simulator and the 5G NR standards, is used to simulate practical FDD cellular systems and generate data for training a deep generative model. In this paper, a CVAE is trained offline, and subsequently, the decoder of the CVAE can be utilized in a practical system to enable the robust beamforming based on stochastic WMMSE.
    }
    \label{fig: system}
\end{figure*}

\section{Digital Twin-Inspired Offline Learning Scheme} \label{sec:digital_twin_inspired_offline_learning}

The development of wireless communication simulation technologies has made it feasible to create a reliable channel simulator tailored to specific wireless communication scenarios  \cite{real-channel-1, real-channel-2, real-channel-4} . Such a simulator has the capability to produce channel samples that closely match the statistical characteristics observed in practical measurements. One recent example is the emergence of the digital twin paradigm \cite{fuller2020digital}. Leveraging the channel simulator, we can offline train a deep generative model, specifically, the CVAE proposed in this paper, to model the channel statistical distribution. The benefit of offline training is that the most computationally demanding training process can be decoupled from the online 5G wireless systems. In other words, we can train a CVAE with huge parameters sufficiently before its online deployment, thus not affecting the operation of practical systems. However, the performance of such a scheme highly relies on the reliability of the adopted channel simulators. In this section, we introduce the design of a digital twin-inspired offline training scheme using the QuaDRiGa channel simulator \cite{real-channel-1}. We also outline the principle of the proposed CVAE and provide a detailed introduction of its structure.

\subsection{Digital Twin-Inspired Offline Learning Scheme}
Before delving into the specifics of the CVAE, we first provide an overview of the proposed CVAE-based offline training method.  The main idea of the proposed method is depicted in Fig. \ref{fig: system} and summarized as follows: Based on the QuaDRiGa channel simulator, we first build a rudimentary ``digital twin" system that can generate the discrete PMI feedback $\{m _n\}_{n=1}^N$, continuous CQI feedback $\{\eta_n\}_{n=1}^N$ and channel vectors $\{ \boldsymbol h_n\}_{n=1}^N$. Particularly, QuaDRiGa is employed to generate channel vectors $\{ \boldsymbol h_n\}_{n=1}^N$, which can be viewed as practical channel measurements. In addition, following 5G NR standards, the process of feeding PMIs $\{ m_n\}_{n=1}^N$ and CQIs $\{\eta_n\}_{n=1}^N$ back in practical 5G NR FDD cellular systems is simulated. Consequently, a large amount of channel vectors $ \boldsymbol{h}_{n} $,  the associated PMIs, and CQIs can be generated via the digital twin system. 

Given the data, $\{m_n, \boldsymbol{h}_{n}, \eta_n \}_{n=1}^N$ ($N$ can be quite large), we can learn a probabilistic mapping $p(\boldsymbol h \mid m, \eta)$. Note that the coarse channel estimate $\hat{\boldsymbol{h}}_n$ is a sufficient statistic of the feedback $m_n$, thus the probabilistic mapping $p(\boldsymbol h \mid \hat{\boldsymbol{h}}, \eta)$ is equivalent to $p(\boldsymbol h \mid m, \eta)$. In this paper, we focus on learning the probabilistic mapping $p(\boldsymbol h \mid \hat{\boldsymbol{h}}, \eta)$ with the CVAE, which is introduced in the following subsection. After the training, as illustrated in Fig. \ref{fig: system}, the decoder of the CVAE will be deployed in practical 5G systems. Taking the PMI and CQI as input, a coarse channel estimator will first output the coarse channel estimate. Then, taking the coarse channel estimate as input, the well-trained decoder can output refined channel samples/estimates,  which further serve as the input of the stochastic WMMSE algorithm for robust beamforming.

\subsection{Channel VAE (CVAE) for Channel Statistical Distribution Modeling}
This section introduces the proposed CVAE tailored for simulating the channel statistical distribution. The primary goal of this CVAE is to learn the underlying probabilistic mapping\footnote{Without further specification, hereafter, we will use $p_{\mathcal{D}}(\boldsymbol{h} \vert \hat{\boldsymbol{h}}, \eta)$ to represent the previous $p(\boldsymbol{h} \vert \hat{\boldsymbol{h}}, \eta)$ to avoid any potential confusion with the parametric probability mapping $p_{\boldsymbol{\theta}}(\boldsymbol{h} \vert \hat{\boldsymbol{h}}, \eta)$ constructed later.}, denoted as $p_{\mathcal{D}}(\boldsymbol{h} \mid \hat{\boldsymbol{h}}, \eta)$.  
Note that this probabilistic mapping could be intuitively seen as trying to refine coarse channel estimations. Consequently, we expect that the channel estimation generated from the underlying probabilistic mapping will be closer to actual ground-truth channel vectors compared to coarse channel estimations.  However, in practice, this underlying probabilistic mapping $p_{\mathcal{D}}(\boldsymbol{h} \mid \hat{\boldsymbol{h}}, \eta)$ is unattainable, only samples from it can be obtained. Therefore, instead of directly learning the underlying probabilistic mapping $p_{\mathcal{D}}(\boldsymbol{h} \mid \hat{\boldsymbol{h}}, \eta)$, we construct a parametric probabilistic mapping $p_{\boldsymbol{\theta}}(\boldsymbol h \mid \hat{\boldsymbol{h}}, \eta)$, parameterized by $\boldsymbol{\theta}$. We anticipate that this parametric probabilistic mapping $p_{\boldsymbol{\theta}}(\boldsymbol h \mid \hat{\boldsymbol{h}}, \eta)$ could approach the underlying probabilistic mapping $p_{\mathcal{D}}(\boldsymbol{h} \mid \hat{\boldsymbol{h}}, \eta)$ in terms of KL divergence, mathematically, 
\begin{align}
    \min_{\boldsymbol{\theta}} D_{KL}\left( p_{\mathcal{D}}(\boldsymbol{h} \mid \hat{\boldsymbol{h}}, \eta) \| p_{\boldsymbol{\theta}}(\boldsymbol h \mid \hat{\boldsymbol{h}}, \eta) \right). 
    \label{eq:KL}
\end{align}
Due to the inaccessibility of the underlying probabilistic mapping $p_{\mathcal{D}}(\boldsymbol{h} \mid \hat{\boldsymbol{h}}, \eta)$, an alternative objective function is given in Identity \ref{theo:alter}. 
\begin{identity}
    \begin{align}
        \min_{\boldsymbol{\theta}} D_{KL}\left( p_{\mathcal{D}}(\boldsymbol{h} \mid \hat{\boldsymbol{h}}, \eta) \| p_{\boldsymbol{\theta}}(\boldsymbol h \mid \hat{\boldsymbol{h}}, \eta)\right) \nonumber \\ 
        =  \max_{\boldsymbol{\theta}} \mathbb{E}_{p_{\mathcal{D}}(\boldsymbol{h} \mid \hat{\boldsymbol{h}}, \eta)} \left[ \log p_{\boldsymbol{\theta}}(\boldsymbol h \mid \hat{\boldsymbol{h}}, \eta) \right]. 
    \end{align}
    \label{theo:alter}
\end{identity}
\begin{proof}
    The KL divergence between the underlying probabilistic mapping $p_{\mathcal{D}}(\boldsymbol{h} \mid \hat{\boldsymbol{h}}, \eta)$ and the parametric probabilistic mapping $p_{\boldsymbol{\theta}}(\boldsymbol h \mid \hat{\boldsymbol{h}}, \eta)$ can be expanded as 
    \begin{align}
        & D_{KL}( p_{\mathcal{D}}(\boldsymbol{h} \mid \hat{\boldsymbol{h}}, \eta) \| p_{\boldsymbol{\theta}}(\boldsymbol h \mid \hat{\boldsymbol{h}}, \eta)) \\ 
        & = - \mathbb{E}_{p_{\mathcal{D}}(\boldsymbol{h} \mid \hat{\boldsymbol{h}}, \eta)} \left[ \log p_{\boldsymbol{\theta}}(\boldsymbol h \mid \hat{\boldsymbol{h}}, \eta)   \right] \nonumber \\
        &~~~~+  \mathbb{E}_{p_{\mathcal{D}}(\boldsymbol{h} \mid \hat{\boldsymbol{h}}, \eta)} \left[ \log p_{\mathcal{D}}(\boldsymbol h \mid \hat{\boldsymbol{h}}, \eta)   \right].
    \end{align}
    Here, the second term on the right-hand side (RHS) is irrelevant to $\boldsymbol{\theta}$, which leads to 
    \begin{align}
        \min_{\boldsymbol{\theta}} D_{KL} \left( p_{\mathcal{D}}(\boldsymbol{h} \mid \hat{\boldsymbol{h}}, \eta) \| p_{\boldsymbol{\theta}}(\boldsymbol h \mid \hat{\boldsymbol{h}}, \eta) \right)  \\ 
        =  \max_{\boldsymbol{\theta}} \mathbb{E}_{p_{\mathcal{D}}(\boldsymbol{h} \mid \hat{\boldsymbol{h}}, \eta)} \left[ \log p_{\boldsymbol{\theta}}(\boldsymbol h \mid \hat{\boldsymbol{h}}, \eta) \right]. \nonumber 
    \end{align}
\end{proof}
Then, the optimization problem in \eqref{eq:KL} could be transferred into 
\begin{align}
     \max_{\boldsymbol{\theta}} \mathbb{E}_{p_{\mathcal{D}}(\boldsymbol{h} \mid \hat{\boldsymbol{h}}, \eta)} \left[ \log p_{\boldsymbol{\theta}}(\boldsymbol h \mid \hat{\boldsymbol{h}}, \eta) \right]. 
    \label{eq:marginal}
\end{align}
In the context of VAEs \cite{vae}, the latent variable $\boldsymbol{z}$ is introduced to capture the essential characteristics of channel vectors and could assist in modeling the desired probabilistic mapping $p_{\boldsymbol{\theta}}(\boldsymbol{h} \mid \hat{\boldsymbol{h}}, \eta)$. With the latent variable $\boldsymbol{z}$, the optimization problem in \eqref{eq:marginal} can be rewritten as
\begin{align}
    \max_{\boldsymbol{\theta}} \ \mathbb{E}_{p_{\mathcal{D}}(\boldsymbol{h} \mid \hat{\boldsymbol{h}}, \eta)} \left[ \log \int p_{\boldsymbol{\theta}}(\boldsymbol h \mid \hat{\boldsymbol{h}}, \eta, \boldsymbol{z}) p(\boldsymbol{z}) d \boldsymbol{z} \right],
    \label{eq:with_z_obj}
\end{align}
where the prior distribution of the latent variable, denoted as $p(\boldsymbol z)$, is typically assumed to be the standard Gaussian distribution. Therefore, it is necessary to learn the probabilistic mapping from the latent variable $\boldsymbol{z}$ to the channel vector $\boldsymbol{h}$, given the coarse channel estimation $\hat{\boldsymbol{h}}$ and CQI $\eta$, i.e.,  $p_{\boldsymbol{\theta}}(\boldsymbol{h} \mid \hat{\boldsymbol{h}}, \boldsymbol{z}, \eta)$, also known as the \textit{decoder} in the VAE literature. Specifically, the decoder is modeled as
\begin{align}
  p_{\boldsymbol{\theta}}(\boldsymbol{h} \mid \hat{\boldsymbol{h}}, \boldsymbol{z}, \eta) = \mathcal{N}(f_{\boldsymbol{\theta}}(\hat{\boldsymbol{h}}, \boldsymbol{z}, \eta), \sigma^2 \boldsymbol{I}),
    \label{decoder} 
\end{align} 
where $\sigma^2$ denotes the variance, which is a hyperparameter and the mean function, $f_{\boldsymbol{\theta}}(\cdot)$, is modeled by a neural network. Even though the decoder has an explicit form, it is still challenging to maximize the objective function in \eqref{eq:with_z_obj} with respect to $\boldsymbol \theta$, due to the intractable integration. To resolve this circumstance, an approximated posterior distribution $q_{\boldsymbol{\phi}}(\boldsymbol{z} \mid \boldsymbol{h}, \hat{\boldsymbol{h}}, \eta)$, parameterized by $\boldsymbol{\phi}$
is introduced, which is also known as the \textit{encoder}. The encoder $q_{\boldsymbol{\phi}}(\boldsymbol{z} \mid \boldsymbol{h}, \hat{\boldsymbol{h}}, \eta)$ is assumed to follow a Gaussian distribution, with the mean $\boldsymbol{\mu}_{\boldsymbol{\phi}}$ and the covariance $\boldsymbol{\Sigma}_{\boldsymbol{\phi}}$ parameterized by neural networks with parameters represented as  $\boldsymbol{\phi}$ \cite{vae}, i.e.,
\begin{align}
  q_{\boldsymbol{\phi}}(\boldsymbol{z} \mid \boldsymbol{h}, \hat{\boldsymbol{h}}, \eta) = \mathcal{N}( \boldsymbol{\mu}_{\boldsymbol{\phi}}, \boldsymbol{\Sigma}_{\boldsymbol{\phi}}), 
  \label{encoder}
\end{align}

With the introduced encoder, the objective function in \eqref{eq:with_z_obj} can be decomposed into two terms:
\begin{align}
     & \mathbb{E}_{p_{\mathcal{D}}(\boldsymbol{h} \mid \hat{\boldsymbol{h}}, \eta)} \left[ \log p_{\boldsymbol \theta}(\boldsymbol h \mid \hat{\boldsymbol{h}}, \eta) \right]   \\ 
    & = \log D_{KL}\left( q_{\boldsymbol{\phi}}(\boldsymbol z \mid \boldsymbol h, \hat{\boldsymbol{h}}, \eta) || p_{\boldsymbol{\theta}}(\boldsymbol z \mid \boldsymbol h,  \hat{\boldsymbol{h}}, \eta)  \right) \nonumber \\ 
    &~~~+ \underbrace{\log \int q_{\boldsymbol{\phi}}(\boldsymbol z \mid \boldsymbol h , \hat{\boldsymbol{h}}, \eta) \log  \frac{p_{\boldsymbol \theta}(\boldsymbol h, \boldsymbol z \mid  \hat{\boldsymbol{h}}, \eta)}{q_{\boldsymbol{\phi}}(\boldsymbol z \mid \boldsymbol h, \hat{\boldsymbol{h}}, \eta)}}_{\mathcal{L}(q)} \mathrm{d} \boldsymbol{z}. 
\end{align}
The first RHS term is the KL divergence between the approximated posterior distribution $q_{\boldsymbol{\phi}}(\boldsymbol z \mid \boldsymbol h, \hat{\boldsymbol{h}}, $ $ \eta)$ and the ground-truth posterior distribution $p_{\boldsymbol{\theta}}(\boldsymbol z \mid \boldsymbol h, \hat{\boldsymbol{h}}, \eta)$. Since this KL divergence is non-negative, the second RHS term $\mathcal{L}(q)$, known as evidence lower bound (ELBO), is a lower bound on objective functions given in \eqref{eq:marginal} and \eqref{eq:with_z_obj}, which can be written as
\begin{align}
    \mathcal{L}(q)
    &= \underbrace{ \mathbb{E}_{ p_{\mathcal{D}}(\boldsymbol h \mid \hat{\boldsymbol{h}}, \eta)} \left[-D_{KL}\left( q_{\boldsymbol{\phi}}(\boldsymbol z \mid \hat{\boldsymbol h}, \boldsymbol h, \eta) || p(\boldsymbol z)  \right) \right]}_{\mathcal{L}_\text{KL}} \nonumber \\
    &~~~+ \underbrace{\mathbb{E}_{p_{\mathcal{D}}(\boldsymbol h \mid \hat{\boldsymbol{h}}, \eta)} \left[ \mathbb{E}_{q_{\boldsymbol{\phi}}(\boldsymbol z \mid \hat{\boldsymbol h}, \boldsymbol h, \eta)} \left[ \log p_{\boldsymbol{\theta}}(\boldsymbol h \mid \hat{\boldsymbol h}, \boldsymbol z, \eta) \right]\right]}_{\mathcal{L}_{\text{recon}}}.  \label{eq:elbo}
\end{align}

\begin{figure}[!t]
    \centering
    \includegraphics[width= 0.7 \linewidth]{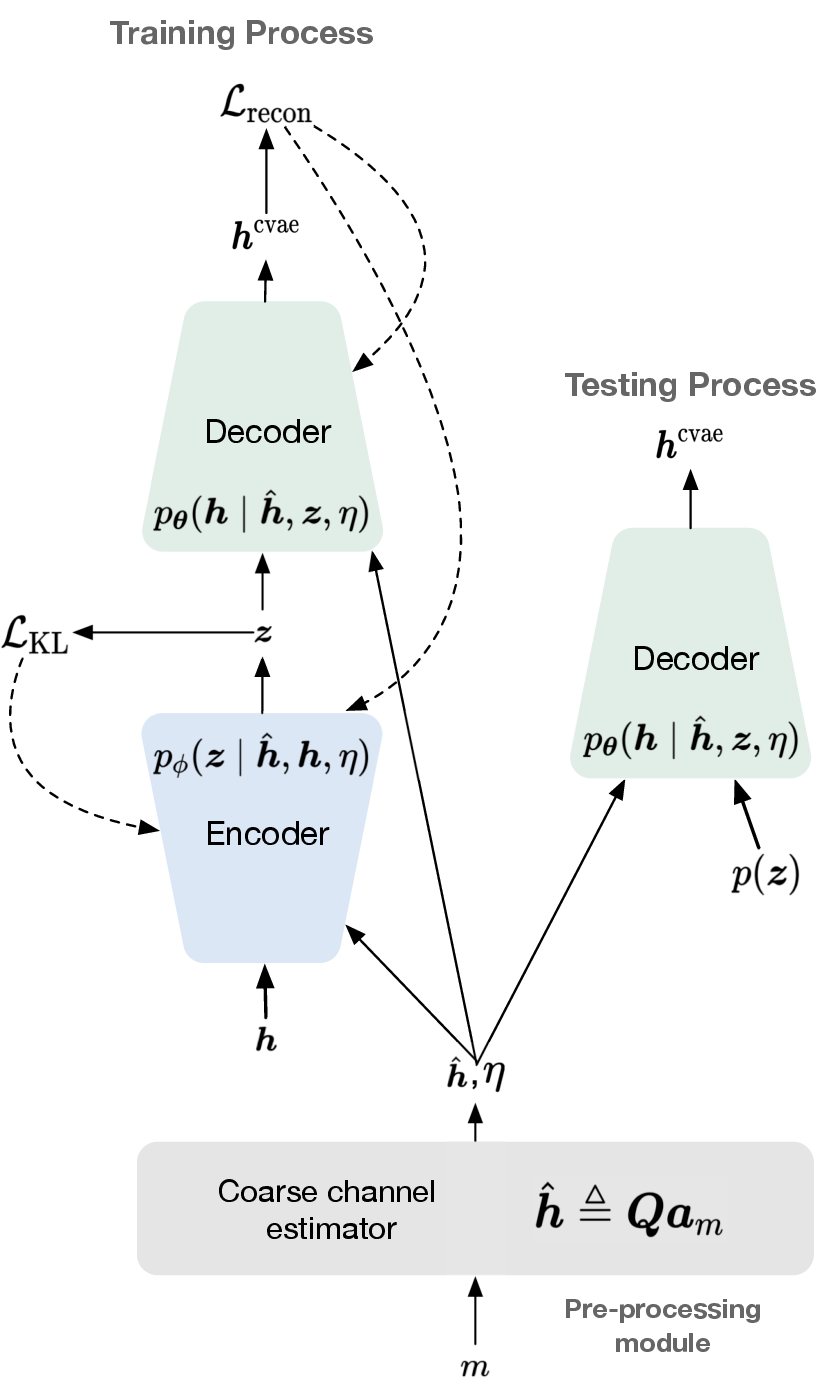}
    \caption{The architecture of the proposed CVAE. Solid lines represent forward flow, and dashed lines represent backward flow.}
    \label{fig:cvae}
\end{figure}
Note that ELBO serves as a lower bound for the objective function defined in \eqref{eq:marginal}. Consequently, maximizing ELBO corresponds to maximizing the objective function in \eqref{eq:marginal}. It is worth recalling that maximizing this objective function is equivalent to minimizing the KL divergence between the underlying probabilistic mapping $p_{\mathcal{D}}(\boldsymbol{h} \mid \hat{\boldsymbol{h}}, \eta)$ and the parametric probabilistic mapping $p_{\boldsymbol{\theta}}(\boldsymbol{h} \mid \hat{\boldsymbol{h}}, \eta)$. 
So, essentially, maximizing ELBO can also be interpreted as an effort to refine coarse channel estimations. This refinement aims to make channel estimations generated by the decoder approach to ground-truth channel vectors more closely than coarse channel estimations.

The term inside the expectation of the reconstruction loss $L_{\text{recon}}$ measures the accuracy of channel refinement and is defined as: 
\begin{align}
     \frac{ | \boldsymbol{h}^H \boldsymbol h^{\text{cvae}} |}{\| \boldsymbol{h} \| \| \boldsymbol h^{\text{cvae}} \|}, 
\end{align} 
which represents the cosine similarity between the ground-truth channel vector $\boldsymbol{h}$ and the output of the decoder $\boldsymbol{h}^{\text{cvae}}$.  

\subsection{Architecture of the Proposed CVAE}
The architecture of the proposed CVAE is explicitly depicted in Fig.~\ref{fig:cvae}. Initially, a coarse channel estimator acts as a pre-processing module to transform the discrete input $m$ into a coarse channel estimate $\hat{\boldsymbol{h}} \triangleq \boldsymbol{Q} \boldsymbol{a}_{m}$, which is then utilized as the input for both the encoder module and the decoder module. As presented on the RHS of Fig.~\ref{fig:cvae}, with the help of the latent variable $\boldsymbol z$ sampled from its prior distribution $p(\boldsymbol z)$, it is expected to learn a decoder that can map the concatenation of the coarse channel estimate $\hat{\boldsymbol h}$ and the CQI $\eta$ to a refined channel estimate $\boldsymbol h^{\text{cvae}}$. To achieve this goal, the CVAE is trained as illustrated on the left-hand side (LHS) of Fig.~\ref{fig:cvae}. It consists of an encoder $p_{\boldsymbol{\phi}}(\boldsymbol{z} \mid \hat{\boldsymbol{h}}, \boldsymbol h, \eta)$ parameterized by $\boldsymbol{\phi}$, and a decoder $p_{\boldsymbol{\theta}}(\boldsymbol h \mid \hat{\boldsymbol{h}}, \boldsymbol{z}, \eta)$ parameterized by $\boldsymbol{\theta}$. Given a set of training data, the encoder takes the concatenation of $\boldsymbol h$, $\hat{\boldsymbol h}$ and the CQI $\eta$ as its input and then outputs a low-dimensional latent variable $\boldsymbol z$, aiming at distilling the information inside the input pair $\{ \boldsymbol h , \hat{\boldsymbol h}, \eta \}$. Then this latent variable $\boldsymbol z$ serves as the input for the decoder to aid the refinement of the coarse channel estimate. The training process is also displayed in Fig.~\ref{fig:cvae}, where solid lines indicate the forward flow, and dash lines represent the backward flow. By minimizing the loss function, in \eqref{eq:elbo}, the decoder and the encoder are alternatively trained using the backpropagation algorithm \cite{li2023overcoming,vae}.

\subsection{Neural Network Design}

\begin{figure}[!t]
    \centering
    \includegraphics[width= \linewidth]{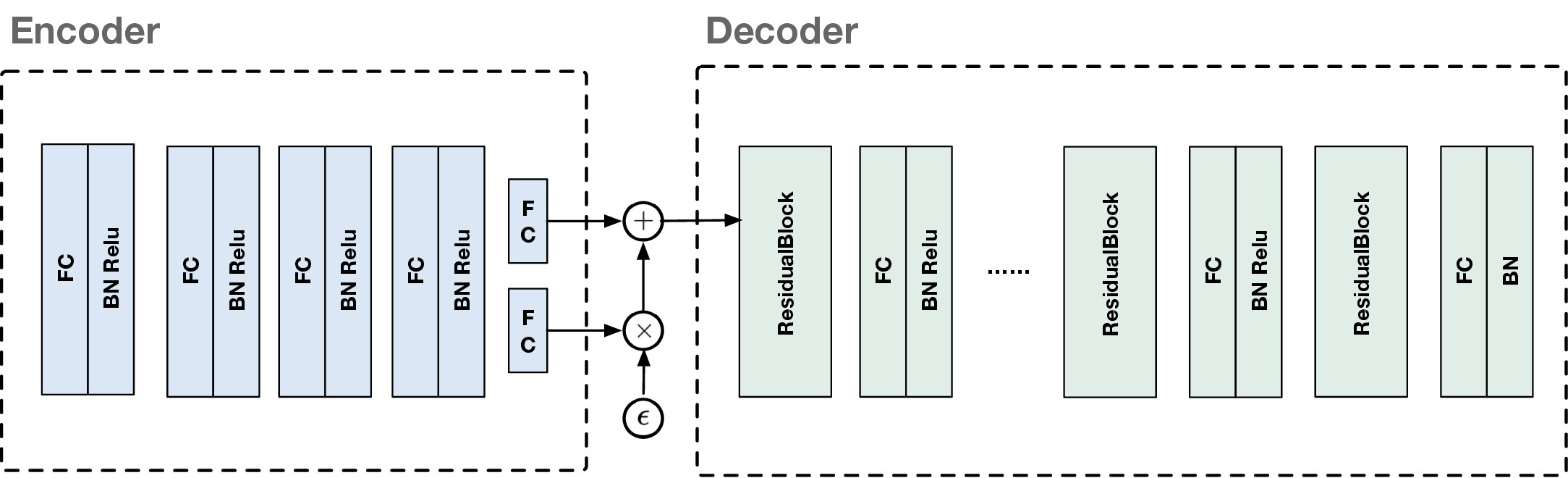}
    \caption{The architecture of the CVAE for the offline scheme consists of: 1. An encoder structure constructed with three so-called FBR units (each FBR unit is composed of a fully connected (FC) layer, a batch normalization (BN) layer, and a ReLU activation function), along with 2 FC layers. 2. A decoder structure with 6 series of residual blocks, 5 FBR units, and an output layer composed of an FC layer and a BN layer.}
    \label{fig:network}
\end{figure}

The neural network utilized in the digital twin-inspired offline learning scheme is presented in Fig. \ref{fig:network} and consists of two essential modules: an encoder and a decoder. The encoder consists of four FBR units followed by two FC layers.
Each FBR unit comprises a fully connected (FC) layer, a batch normalization (BN) layer \cite{theodoridis2020machine}, and a ReLU activation function \cite{theodoridis2020machine}. On the other hand, the decoder is composed of six series of residual blocks \cite{he2016deep}, five FBR units, and an output layer constructed with an FC layer and a BN layer. Each residual block is built with one FBR unit, one FC layer, and one BN layer. The dimensionality of the latent variable $\boldsymbol{z}$ is fixed at $2$.

\section{Type II Codebook-Assisted Online Learning Scheme}  \label{sec:online_learning}

\begin{figure*}[t]
    \centering
    \includegraphics[width= 0.8 \linewidth]{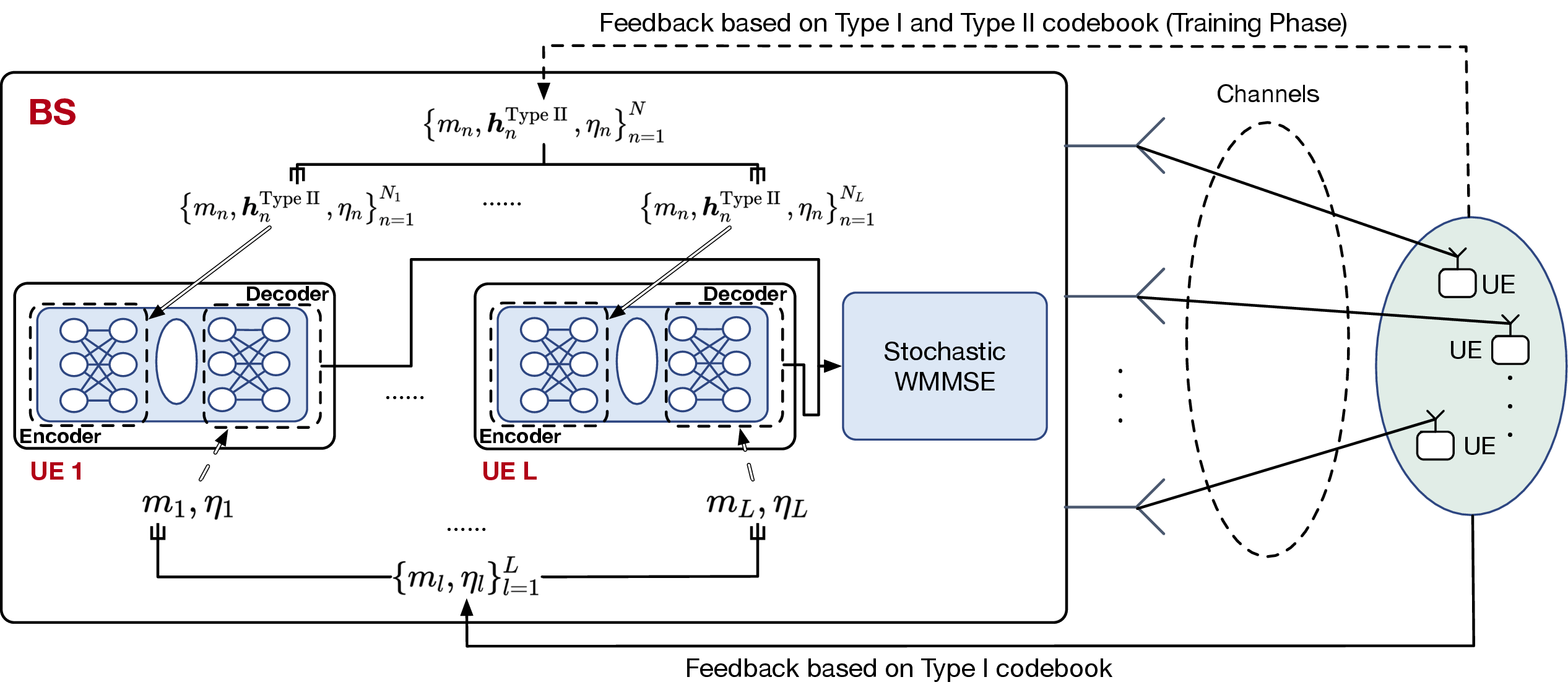}
    \caption{The proposed Type II codebook-assisted online learning scheme. For each UE, a CVAE is trained continuously based on the feedback based on Type I and Type II codebook. 
    Then, the decoder is deployed to refine coarse channel estimates based on Type I codebook, which further 
    to enable the robust beamforming based on the stochastic WMMSE algorithm. Note that bold arrows represent the training process, and bold dotted arrows represent the inference process. } 
    \label{fig:UE_scheme}
\end{figure*}

As discussed in the previous section, the performance of the digital twin-inspired offline learning scheme depends heavily on the reliability of the adopted channel simulator. If channel samples generated by the channel simulator exhibit errors when compared to practical measurements, the learned CVAE may demonstrate poorer generalization performance. To circumvent such a drawback, we propose a Type II codebook-assisted online learning scheme. Particularly, 5G NR standards introduce Type II codebook \cite{liu2016impact}, which results in channel estimations 
\begin{align}
    \left\{ \boldsymbol{h}_n^{\text{Type II}} \triangleq \boldsymbol{Q} \boldsymbol{a}^{\text{Type II}}_{m_n} \right\}_{n=1}^{N}  \nonumber
\end{align}
that are closer to practical channel measurements $\left\{ \boldsymbol{h}_n \right\}_{n=1}^{N}$ than channel estimations based on Type I codebook. However, due to the higher overhead, Type II codebook is not as widely used as Type I codebook in practical wireless systems. However, it can provide accurate estimates of practical channel vectors in the initial  communication rounds. This capability enables us to perform online training for the CVAE for each UE, as explained in the following subsections. 

\subsection{Type II Codebook-Assisted Online Learning Scheme}

The process of the Type II codebook-assisted online learning scheme is depicted in Fig. \ref{fig:UE_scheme}. To balance the trade-off between transmission overhead and estimation accuracy, we assume that both Type II codebook-based feedback and Type I codebook-based feedback can be acquired in the initial communication rounds. After a certain number of communication rounds, the training data set $\left\{m_{n}, \boldsymbol{h}_{n}^{\text {Type II }}, \eta_{n}\right\}_{n=1}^{N}$ can be constructed at the BS. Since channel vectors may vary significantly among different UEs, we individually train a CVAE for each UE using the partitioned training data set $ \left\{m_{n}, \boldsymbol{h}_{n}^{\text {Type II }}, \eta_n \right\}_{n=1}^{N_{l}}$, where $N_l$ represents the number of channel vectors for UE $l$. During the inference phase, as illustrated in Fig. \ref{fig:UE_scheme}, the decoder is deployed. Using PMI $m_l$ and CQI $\eta_l$ as inputs, the coarse channel estimator transforms them into the coarse channel estimate $\hat{\boldsymbol{h}}_l$. Subsequently, with coarse channel estimates as inputs, the decoder for UE $l$ separately outputs the refined channel samples/estimates, enabling the use of the stochastic WMMSE algorithm for robust beamforming. 

Note that in the online learning scheme, Type II codebook-based channel estimates $ \{ \boldsymbol{h}_{n}^{\text {Type II }} \}$ are served for the training purpose, and a CVAE is trained for each UE separately. Therefore, it is expected that the online learning scheme is more adaptable to the fast-changing wireless environments and provides tailed robust beamforming for each UE.

\begin{figure}[t!]
    \centering
    \includegraphics[width= \linewidth]{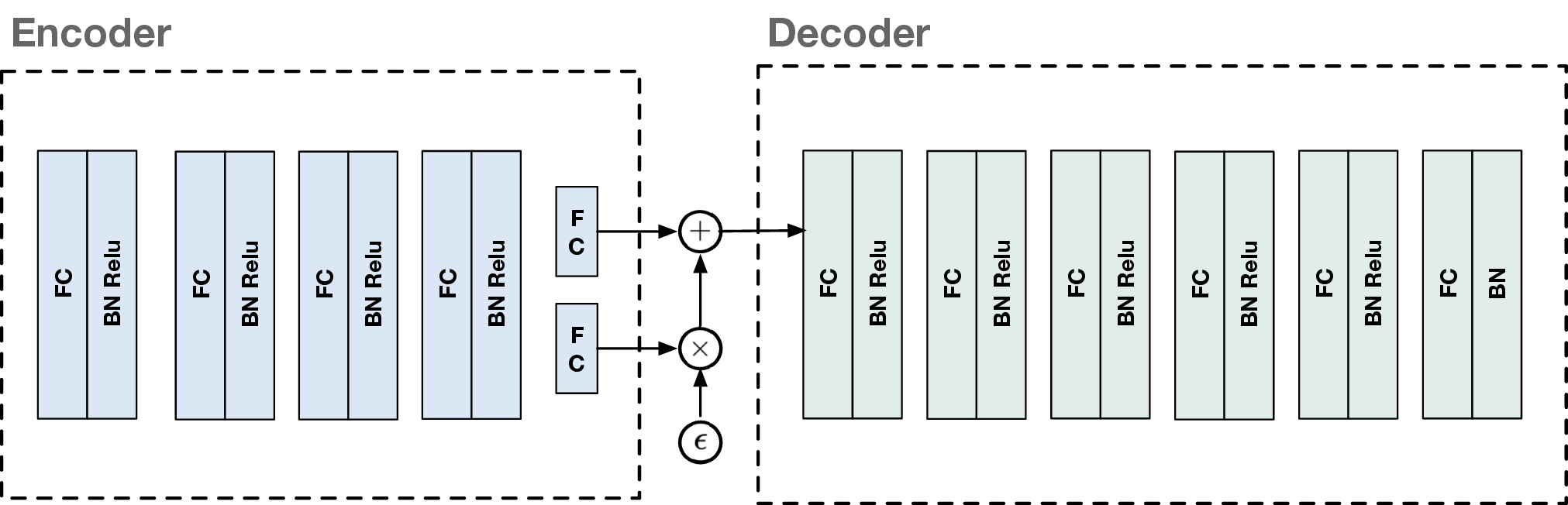}
    \caption{The architecture of the CVAE for the online scheme consists of an encoder structure constructed with four FBR units and two FC layers, and a decoder structure comprising 5 series FBR units and an output layer composed of an FC layer and a BN layer.}
    \label{fig:UEnetwork}
\end{figure}

\begin{remark}
Importantly, in the context of two proposed approaches, namely the online/offline scheme, a trade-off exists between simulation errors and channel estimation errors. Simulation errors arise from errors between the channel simulator and the practical wireless environment, while Type II feedback-based channel estimations lead to channel estimation errors. In the digital twin-inspired offline learning approach, simulation errors are prevalent, and the channel estimation error is non-existent, as it treats the simulated channel as true channel vectors. In contrast, in the Type II codebook-assisted online learning scheme, even with high resolution, channel estimations using Type II codebook deviates from practical channel measurements, degrading performance. Furthermore, the training data for the Type II codebook-assisted online scheme is collected in a practical communication environment, resulting in zero simulation error. Thus, the superiority of one method over the other hinges on which type of error exerts a more substantial impact. If simulation errors have more significant consequences, the online method is expected to yield superior results, whereas the reverse is true if the other type of error carries more weight. To determine the magnitude of these impacts, we will conduct numerical simulations in the following section.
\end{remark}

\subsection{Neural Network Design}

Due to the higher correlation among channel vectors of a specific UE, the results in the training process imply that a simple network leads to better performance. The neural network we proposed in this paper for the online scheme is shown in Fig. \ref{fig:UEnetwork} and is divided into two parts: the encoder and the decoder. The encoder part consists of four FBR units and $2$ FC layers. The decoder part consists of $5$ series FBR units and an output layer, which is composed of an FC layer and a BN layer. The dimension of the latent variable $\boldsymbol z$ is 2. 

\begin{figure*}[t]
    \centering 
    \includegraphics[width= 0.8 \linewidth]{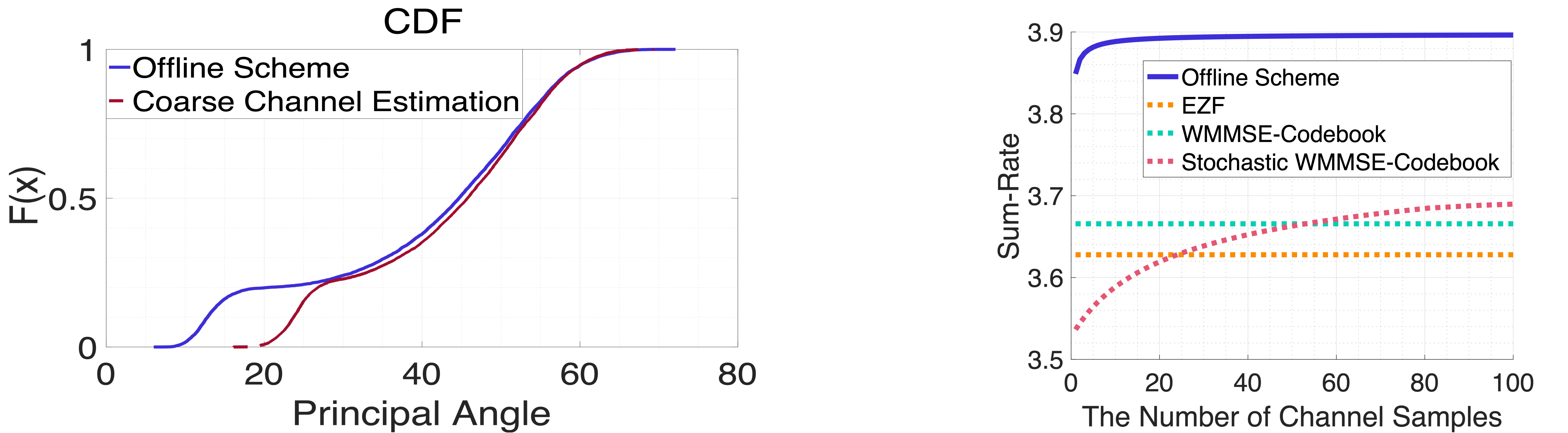}
    \caption{(Left) The cumulative distribution function of generated channel vectors of the offline scheme and coarse
 channel estimates $\{ \hat{\boldsymbol{h}}_i \}_{i=1}^{N_{\text{test}}}$ versus the principal angle. (Right) The sum-rates from different robust beamforming schemes versus the number of channel samples.}
    \label{fig:offline}
\end{figure*} 

\begin{figure*}[t]
    \centering 
    \includegraphics[width= 0.8 \linewidth]{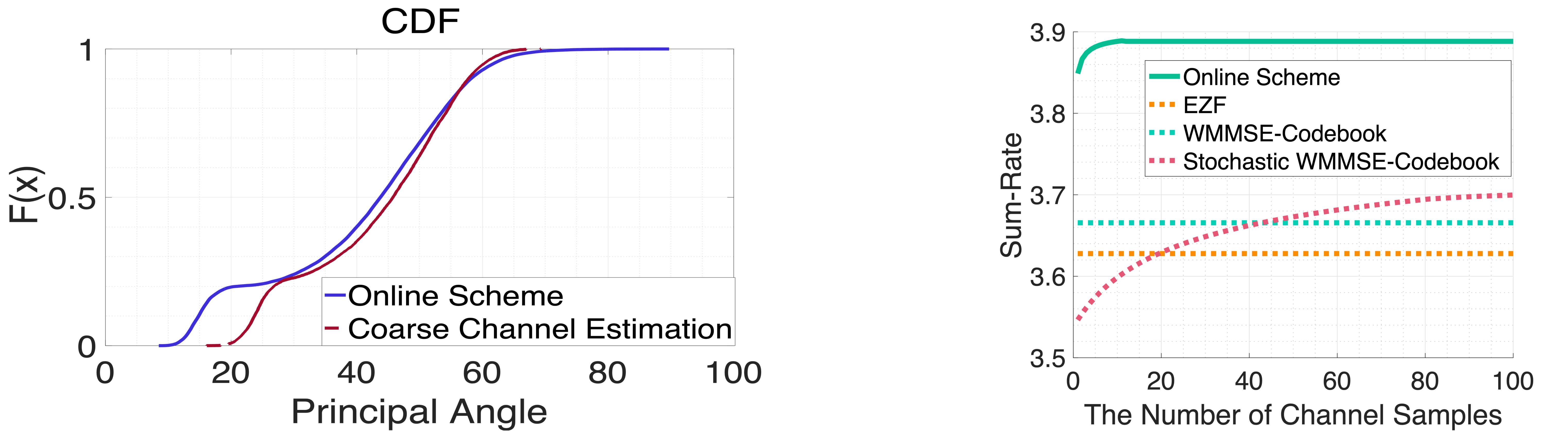}
    \caption{(Left) The cumulative distribution function of generated channel vectors of the online scheme and coarse
channel estimates $\{ \hat{\boldsymbol{h}}_i \}_{i=1}^{N_{\text{test}}}$ versus the principal angle. (Right) The sum-rates from different robust beamforming schemes versus the number of channel samples.}
    \label{fig:online}
\end{figure*}

\section{Numerical Results and Discussions} \label{sec:experiments}

Section \ref{subsec:set_up} provides an overview of the simulation setup. Section \ref{subsec:benchmarks} introduces performance measures and benchmarks utilized in this paper. Finally, simulation results and corresponding discussions are presented in Section \ref{subsec:results}.

\subsection{Simulation Setup}
\label{subsec:set_up}

Consider a BS with $N_A = 32$ antennas and $N_P = 8$ ports. Channel vectors were obtained by QuaDRiGa, where the speed of each UE is $30$ km/h and the sample interval is $5$ ms. Channel vectors of $10$ UEs are generated with a single subcarrier and $10000$ time sampling points. Training sets and test sets contain $N_{\text{train}}=90000$, and $N_{\text{test}}=10000$ channel vectors (including all UEs), respectively. Type I codebook \cite{code1} and Type II codebook \cite{liu2016impact} are adopted and PMIs and CQIs are acquired following 5G NR standards \cite{5g-1, 5g-2}. In summary, the training data set for the online scheme is $\mathcal D_{\text{train}}^{\text{online}} \triangleq \{\boldsymbol h^{\text{Type II}}_n, m_n, \eta_n \}_{n=1}^{N_{\text{train}}}$, that for the offline scheme is $\mathcal D_{\text{train}}^{\text{offline}} \triangleq \{ \boldsymbol{h}_n, m_n, \eta \}_{n=1}^{N_{\text{train}}}$, and the test data set for the online scheme and the offline scheme is $\mathcal D_{\text{test}} \triangleq \{\boldsymbol{h}_n, \boldsymbol h^{\text{Type II}}_n, m_n, \eta_n \}_{n=1}^{N_{\text{test}}}$.  For all experiments, we use the Adam optimizer \cite{adam} for training with a learning rate $0.001$ and rescale it with $0.1$ at epoch $3$, $7$. The training epoch is $10$ with batch size $64$. 

\subsection{Performance Measure and Benchmarks}
\label{subsec:benchmarks}
To assess the beamforming performance, we employ two measures: the sum-rate (defined in the objective of \eqref{ssum_rate}) and the principal angle. Here the principal angle, defined as
\begin{align}
    \operatorname{arcos}\left(\frac{|\boldsymbol{h}^{H} \hat{\boldsymbol{h}}|}{\|\boldsymbol{h}\|\|\hat{\boldsymbol{h}}\|}\right), 
\end{align}
serves as a similarity metric between the estimated channel vector $\hat{\boldsymbol{h}}$ and the ground-truth channel vector $\boldsymbol{h}$. To establish a performance baseline, we compare our proposed methods against the extended zero-forcing (EZF) \cite{ezf} and WMMSE \cite{wmmse} algorithms by reporting their corresponding sum-rates.



\begin{figure*}[t]
    \centering 
    \includegraphics[width= 0.8 \linewidth]{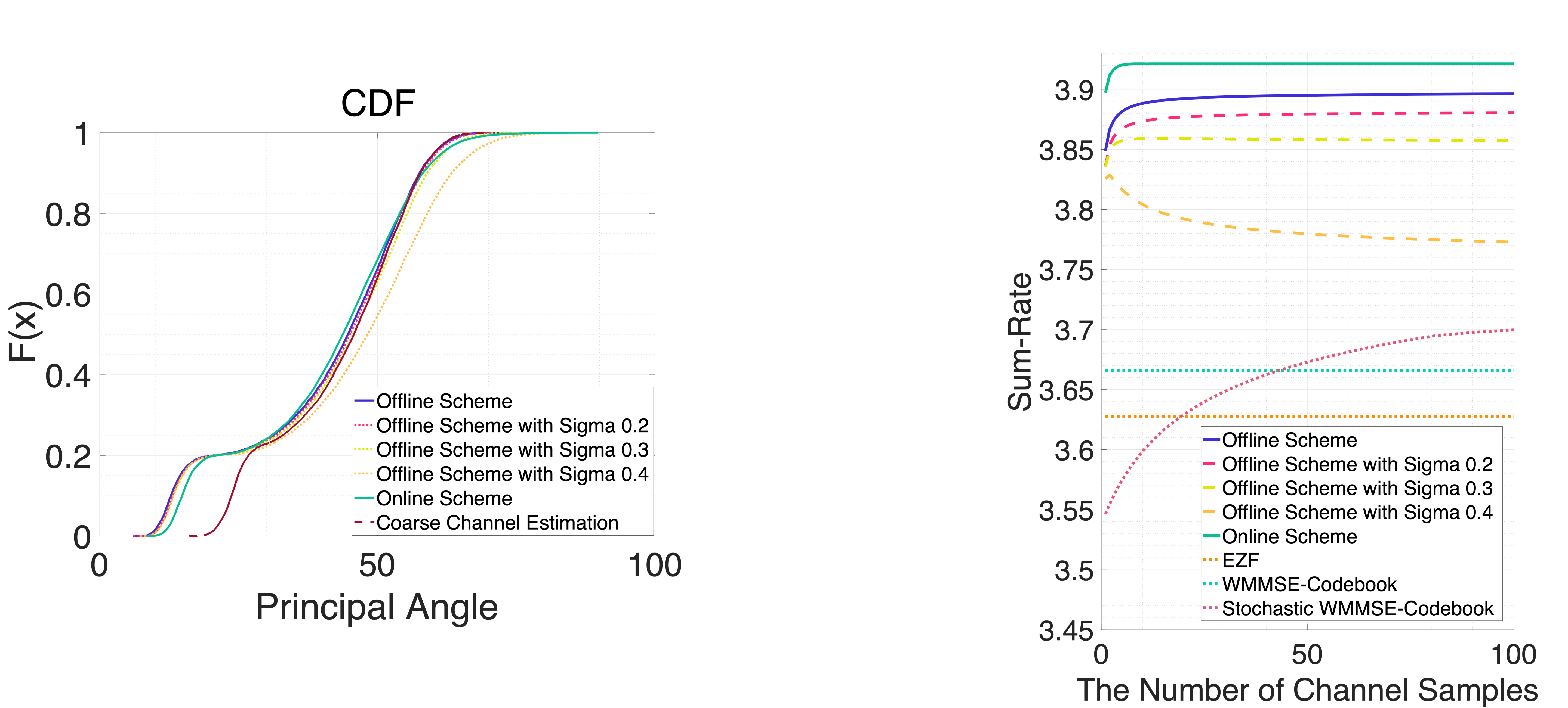}
    \caption{(Left) The cumulative distribution function of generated channel vectors of different schemes and coarse
channel estimates $\{ \hat{\boldsymbol{h}}_i \}_{i=1}^{N_{\text{test}}}$ versus the principal angle. (Right) The sum-rates from different robust beamforming schemes versus the number of channel samples.}
    \label{fig:online_vs_offline}
\end{figure*} 

\subsection{Simulation Results}
\label{subsec:results}

\subsubsection{Offline Scheme Simulation Results}

To show the accuracy of the proposed CVAE-based channel generator of the offline scheme, 
the cumulative distribution functions (CDFs) of generated channel vectors based on the offline scheme and coarse channel estimates $\{ \hat{\boldsymbol{h}}_i \}_{i=1}^{N_{\text{test}}}$ in terms of the principal angle (evaluated on the test data set) are shown in the LHS of Fig. \ref{fig:offline}. It is clear that given a principal angle smaller than $58^\circ$, the number of channel samples of the offline scheme is always larger than the number of coarse channel estimates $\{ \hat{\boldsymbol{h}}_i \}_{i=1}^{N_{\text{test}}}$. This implies that generated channel samples of the offline scheme is much closer to ground-truth channel vectors.

Then, given the same PMI feedback, the sum-rate of stochastic WMMSE \cite{ssum} based on $100$ samples generated by the proposed CVAE-based channel generator of the offline scheme (denoted as offline scheme) and the sum-rate of stochastic WMMSE based on $100$ samples generated from the coarse channel distribution (denoted as stochastic WMMSE-codebook) are shown in the RHS of Fig. \ref{fig:offline}. From Fig. \ref{fig:offline}, it can be seen that the performance of stochastic WMMSE with samples generated from the proposed CVAE-based channel generator of the offline scheme is better than those of the other methods. It is shown that the proposed CVAE-based channel generator of the offline scheme helps the stochastic WMMSE algorithm achieve enhanced sum-rate performance.



\subsubsection{Online Scheme Simulation Results}

To demonstrate the accuracy of the proposed CVAE-based channel generator of the online scheme, we present CDFs of generated channel vectors based on the online scheme and coarse channel estimates $\{ \hat{\boldsymbol{h}}_i \}_{i=1}^{N_{\text{test}}}$ in terms of the principal angle in the LHS of Fig. \ref{fig:online}. As shown, for principal angles smaller than $55^\circ$, the number of channel samples of the online scheme is consistently higher than the number of coarse channel estimates $\{ \hat{\boldsymbol{h}}_i \}_{i=1}^{N_{\text{test}}}$, indicating that generated channel samples of the online scheme are much closer to ground-truth channel vectors.

Then, given the same PMI feedback, the sum-rate of stochastic WMMSE \cite{ssum} based on $100$ samples generated by the proposed CVAE-based channel generator of the online scheme (denoted as online scheme) and the sum-rate of stochastic WMMSE based on $100$ samples generated from the coarse channel distribution are shown in the RHS of Fig. \ref{fig:online}.  This figure clearly shows that the performance of stochastic WMMSE with samples generated from the proposed CVAE-based channel generator of the online scheme is better than that of the other methods even though treating Type-II codebook-based channel estimations as ground-truth channel vectors. This indicates that channel estimation errors have little impact on the performance.  It also confirms that the proposed CVAE-based channel generator of the online scheme assists in enhancing the sum-rate performance of the stochastic WMMSE algorithm. 

\begin{table}[ht]
\centering
\caption{Maximum of the sum-rate versus the number of  time sampling points for each UE. Note that, the sum-rate based on the WMMSE algorithm is $3.6656$.}
\begin{tabular}{r|c|c|c|c|c}
\toprule
\begin{tabular}[r]{@{}r@{}} The number of  \\ time   sampling   \\ points  for each UE \end{tabular}   & 100 & 500 & 1000 & 5000 & 9000 \\ \midrule
\begin{tabular}[r]{@{}r@{}} Maximum of Sum-rate \end{tabular}                                                                     &  3.1818    &  3.4844 &  3.7097    & 3.8041     & 3.9213 \\ \bottomrule
\end{tabular}
\label{tab:1}
\end{table}

To further demonstrate the performance of the online scheme, the relationship between the maximum sum-rate of the online scheme on the test data set and the number of time sampling points in each UE in the training data set is shown in Table \ref{tab:1}. Note that the test data set is fixed and contains the last $1000$ time sampling points of each UE. It can be seen that as the number of time sampling points for each UE increases, the maximum value of sum-rate continues to rise. Specifically, when the number of time sampling points per UE is equal to $1000$, the performance of the online scheme outperforms that of WMMSE with coarse channel estimations. This indicates that the online scheme can effectively capture the time correlation between CSI even just using $1/10$ time sampling points for each UE. This further demonstrates the superiority of the online scheme, that is, for each UE, we only need to collect a small part of training data to obtain a relatively good channel statistical distribution for unseen samples.

\subsubsection{Online Scheme Versus Offline Scheme}

To illustrate the impact of simulation errors, we introduced noise following AWGN with varying variances into channel vectors of training data sets used in the offline scheme. Additionally, we labeled the offline scheme with variance 0.2 as \textit{Offline Scheme with Sigma 0.2}. Schemes with other variances use similar naming conventions. Generated channel vectors of different schemes and coarse channel estimates $\{ \hat{\boldsymbol{h}}_i \}_{i=1}^{N_{\text{test}}}$ are presented in terms of the principal angle CDFs in the LHS of Fig. \ref{fig:online_vs_offline}. As the variance of noise increases, the CDFs of the offline scheme move towards the right side. This indicates that generated channel samples gradually deviate from ground-truth channel vectors as the variance of noise increases.

After receiving the same PMI feedback, the sum-rates of stochastic WMMSE \cite{ssum} were calculated using $100$ samples generated from the proposed CVAE-based channel generator of different schemes, and results are presented in the RHS of Fig. \ref{fig:online_vs_offline}. As shown in this figure, stochastic WMMSE with samples generated from the proposed CVAE-based channel generator of the online scheme performs better than the other methods, indicating that training a UE with an individual CVAE can improve performance despite channel estimation errors. Moreover, as the variance of the noise increases, the performance of the proposed CVAE-based channel generator of the offline scheme decreases, although it still outperforms benchmark methods. This raises a trade-off between performance gains and operation complexity, where the online scheme offers better performance but may impact the operation of practical systems. In contrast, the performance of the offline scheme may decrease due to simulation errors but can be decoupled from the online 5G wireless system.

\section{Conclusion}
\label{sec:conclusion}

This paper introduces the CVAE system, tailored to simulate the channel statistical distribution without relying on specific distribution assumptions. To enable seamless integration of model learning in practical wireless communication environments, we introduce two comprehensive learning strategies for preparing the CVAE model for practical deployment. The first strategy involves offline training using a high-performance channel simulator (e.g., QuaDRiGa), which generates extensive training data closely mirroring authentic channel characteristics. The second method, applicable when a high-performance simulator is unavailable, introduces a Type II codebook-assisted online learning scheme, allowing adaptive responses to evolving wireless environments and ensuring robust beamforming for individual UE. The channel statistical distributions resulting from these approaches enable the practical implementation of the stochastic WMMSE technique in practical 5G NR FDD cellular systems. Numerical results provided in the paper illustrate the effectiveness of CVAE and the associated learning methods in enhancing robust beamforming performance.

\bibliographystyle{IEEEtran}
\bibliography{main}{}


 





\end{document}